\documentclass[aps,pra,reprint,superscriptaddress,nofootinbib]{revtex4-2}

\usepackage{graphicx} 
\usepackage{amsmath}
\usepackage{amssymb}
\usepackage{hyperref}
\usepackage{float}
\usepackage{lipsum} 

\begin{document}

\title{Probing and Enhancing the Robustness of GNN-based QEC Decoders with Reinforcement Learning}

\author{Ryota Ikeda}
\affiliation{Department of Electrical and Electric Engineering, Yamaguchi University}

\date{\today}

\begin{abstract}
Graph Neural Networks (GNNs) have emerged as a powerful, data-driven approach for Quantum Error Correction (QEC) decoding, capable of learning complex noise characteristics directly from syndrome data. However, the robustness of these decoders against subtle, adversarial perturbations remains a critical open question. This work introduces a novel framework to systematically probe the vulnerabilities of a GNN decoder using a reinforcement learning (RL) agent. The RL agent is trained as an adversary with the goal of finding minimal syndrome modifications that cause the decoder to misclassify. We apply this framework to a Graph Attention Network (GAT) decoder trained on experimental surface code data from Google Quantum AI. Our results show that the RL agent can successfully identify specific, critical vulnerabilities, achieving a high attack success rate with a minimal number of bit flips. Furthermore, we demonstrate that the decoder's robustness can be significantly enhanced through adversarial training, where the model is retrained on the adversarial examples generated by the RL agent. This iterative process of automated vulnerability discovery and targeted retraining presents a promising methodology for developing more reliable and robust neural network decoders for fault-tolerant quantum computing.
\end{abstract}

\maketitle

\section{Introduction}

The realization of fault-tolerant quantum computers hinges on the efficacy of Quantum Error Correction (QEC) \cite{shor1995scheme}. A crucial component of any QEC scheme is the decoder, a classical algorithm tasked with inferring the most likely physical errors from a set of syndrome measurements. While traditional decoders like Minimum-Weight Perfect Matching (MWPM) \cite{dennis2002topological} perform optimally under simplified noise models, their performance can degrade when faced with the complex, correlated noise present in real quantum devices \cite{google2023suppressing}.

Machine learning, particularly Graph Neural Networks (GNNs), offers a promising alternative by learning the noise model directly from experimental data \cite{torlai2017neural, chamberland2023graph}. Our previous work demonstrated the effectiveness of a Graph Attention Network (GAT)-based decoder that processes syndrome data by "flattening" the time dimension into node features, thereby capturing spatio-temporal correlations without relying on prior theoretical knowledge from models like MWPM \cite{ikeda2025pi_gnn}.

While these data-driven decoders show high accuracy, their reliability and robustness are not well understood. Neural networks are known to be susceptible to adversarial attacks, where small, often imperceptible, perturbations to the input can cause drastic changes in the output \cite{goodfellow2014explaining}. In the context of QEC, such a vulnerability could be catastrophic, leading to undetected logical errors. This raises a critical question: Do GNN-based decoders possess inherent "blind spots" or vulnerabilities that could be systematically exploited?

In this paper, we propose a framework to answer this question by employing a Reinforcement Learning (RL) agent as an automated adversary. The agent's task is to interact with a fixed, pre-trained GAT decoder (the "environment") and learn a policy for flipping the minimum number of syndrome bits required to induce a misclassification. We then use the adversarial examples discovered by this agent to retrain the original decoder, a technique known as adversarial training, to enhance its robustness. Our contributions are threefold: (1) We introduce an RL-based methodology for systematically probing the vulnerabilities of GNN decoders. (2) We demonstrate its effectiveness by identifying critical weak points in a GAT decoder. (3) We show that adversarial training using these discovered vulnerabilities significantly improves the decoder's robustness against such attacks, illustrating a clear path toward building more reliable decoders.

\section{Methodology}

\subsection{Dataset and Graph Representation}
We use the public surface code experimental dataset from Google Quantum AI \cite{google2023suppressing}, which consists of 50,000 shots. The raw measurement data was processed using the Stim package \cite{gidney2021stim} to generate consistent pairs of syndrome data (\verb|detection_events|) and the corresponding logical flip labels (\verb|obs_flips|).

To represent the spatio-temporal syndrome data as a static graph suitable for GNN processing, we employ a "time-flattening" approach. For a code with $N_s$ spatial detector locations and $T$ measurement rounds, we construct a graph with $N_s$ nodes. The feature vector for each node is a $T$-dimensional vector representing the syndrome outcomes at that spatial location across all rounds. For our dataset, this results in graphs with 4 nodes, each having a 2-dimensional feature vector. The graph is fully connected to ensure information propagation between all spatial nodes.

\subsection{Target Decoder: GAT-GNN Architecture}
The target of our adversarial attack is the GAT-based decoder from our prior work \cite{ikeda2025pi_gnn}. The model consists of two \verb|GATv2Conv| layers \cite{brody2021how}, each followed by \verb|LayerNorm| and a `ReLU` activation. A \verb|global_mean_pool| layer aggregates node features into a single graph-level representation, which is then passed to a two-layer MLP classifier to predict the probability of a logical error.

The model is trained to minimize a weighted binary cross-entropy loss. This addresses the severe class imbalance in the dataset (46,030 negative vs. 3,970 positive samples). The loss function is:
\begin{equation}
    L_{\text{data}} = \text{BCEWithLogitsLoss}(y_{\text{pred}}, y_{\text{true}}, \text{pos\_weight}=w_p)
\end{equation}
where the weight $w_p = N_{\text{neg}} / N_{\text{pos}}$ is assigned to the positive class.

\subsection{Adversarial Agent: RL-based GAT Actor}
We frame the task of finding adversarial examples as an RL problem.

\begin{itemize}
    \item \textbf{Environment}: The pre-trained and fixed GAT-GNN decoder. It receives a state (syndrome graph) and returns the predicted logical error probability.
    \item \textbf{State ($s_t$)}: The syndrome graph at step $t$, represented by a \verb|torch_geometric.data.Data| object.
    \item \textbf{Action ($a_t$)}: An integer representing the index of the syndrome bit to flip. The action space size is $N_s \times T$.
    \item \textbf{Reward ($r_t$)}: The goal is to flip a syndrome from being classified as "no error" to "error". The reward is defined as the change in the decoder's predicted logical error probability, $P_L$. For a state transition from $s_t$ to $s_{t+1}$ via action $a_t$, the reward is:
    \begin{equation}
        r_t = P_L(s_{t+1}) - P_L(s_t)
    \end{equation}
    where $P_L(s) = \sigma(\text{GAT\_GNN}(s))$. This encourages the agent to take actions that maximally increase the decoder's error probability.
\end{itemize}

The RL agent, or "Actor," is itself a GNN with a similar GAT architecture but without the global pooling layer. It processes the input graph and outputs a probability distribution over the entire action space using a policy head and a softmax activation.

\subsection{Training and Evaluation Protocol}
The Actor is trained using the REINFORCE algorithm. An episode begins by sampling a syndrome graph that the target GAT decoder correctly classifies as having no logical error. The Actor then sequentially flips bits for a maximum number of steps, or until the decoder's prediction flips to $P_L > 0.5$. At the end of each episode, the Actor's weights are updated using the collected trajectory of states, actions, and rewards to maximize the expected discounted return $G_t = \sum_{k=0}^{\infty} \gamma^k r_{t+k}$. The policy loss is given by:
\begin{equation}
    L_{\text{policy}} = -\sum_{t} \log \pi(a_t|s_t) G_t
\end{equation}
After training the RL agent, we perform adversarial training on the GAT decoder. We use the trained Actor to generate adversarial examples from the training set. These examples, paired with their original, correct labels (i.e., "no logical error"), are added to the training data. The GAT decoder is then retrained to minimize a combined loss function:
\begin{equation}
    L_{\text{robust}} = L_{\text{clean}} + \alpha L_{\text{adv}}
\end{equation}
where $L_{\text{clean}}$ is the loss on the original data, $L_{\text{adv}}$ is the loss on the generated adversarial examples, and $\alpha$ is a hyperparameter balancing the two terms.

\section{Results}

\subsection{Probing the Initial Decoder}
We first trained the RL agent for 4,000 episodes against the baseline GAT decoder. The agent quickly learned an effective attack strategy. We evaluated the trained agent on a held-out test set of 9,206 samples, all correctly classified as negative (no logical error) by the baseline decoder. The agent achieved an \textbf{Attack Success Rate (ASR) of 91.2\%}, defined as the percentage of samples it could manipulate to be misclassified as positive ($P_L > 0.5$). Remarkably, the average number of bit flips required for a successful attack was only \textbf{1.02}.

\begin{figure}[H]
    \centering
    \includegraphics[width=0.9\columnwidth]{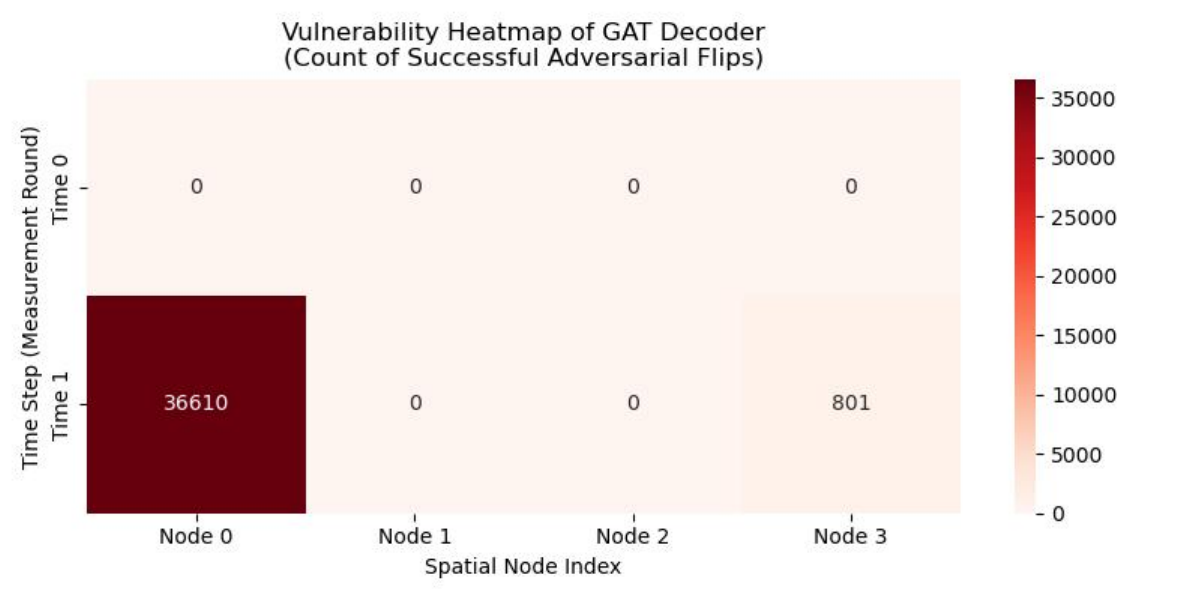}
    \caption{Vulnerability heatmap of the baseline GAT decoder. The color intensity and number indicate the total count of successful attacks initiated by flipping the corresponding syndrome bit. The agent overwhelmingly identified the bit at (Node 0, Time 1) as the most critical vulnerability.}
    \label{fig:heatmap1}
\end{figure}

To understand the agent's learned policy, we aggregated the locations of successful bit flips across all test samples. The resulting vulnerability heatmap is shown in Fig.~\ref{fig:heatmap1}.

The heatmap reveals a stunningly specific vulnerability: 36,610 successful attacks were achieved by flipping the syndrome bit at spatial location `Node 0` in the final measurement round, `Time 1`. This single point represents the decoder's primary "Achilles' heel."

\subsection{Evaluating the Robust Decoder}
\begin{figure}[H]
    \centering
    \includegraphics[width=\columnwidth]{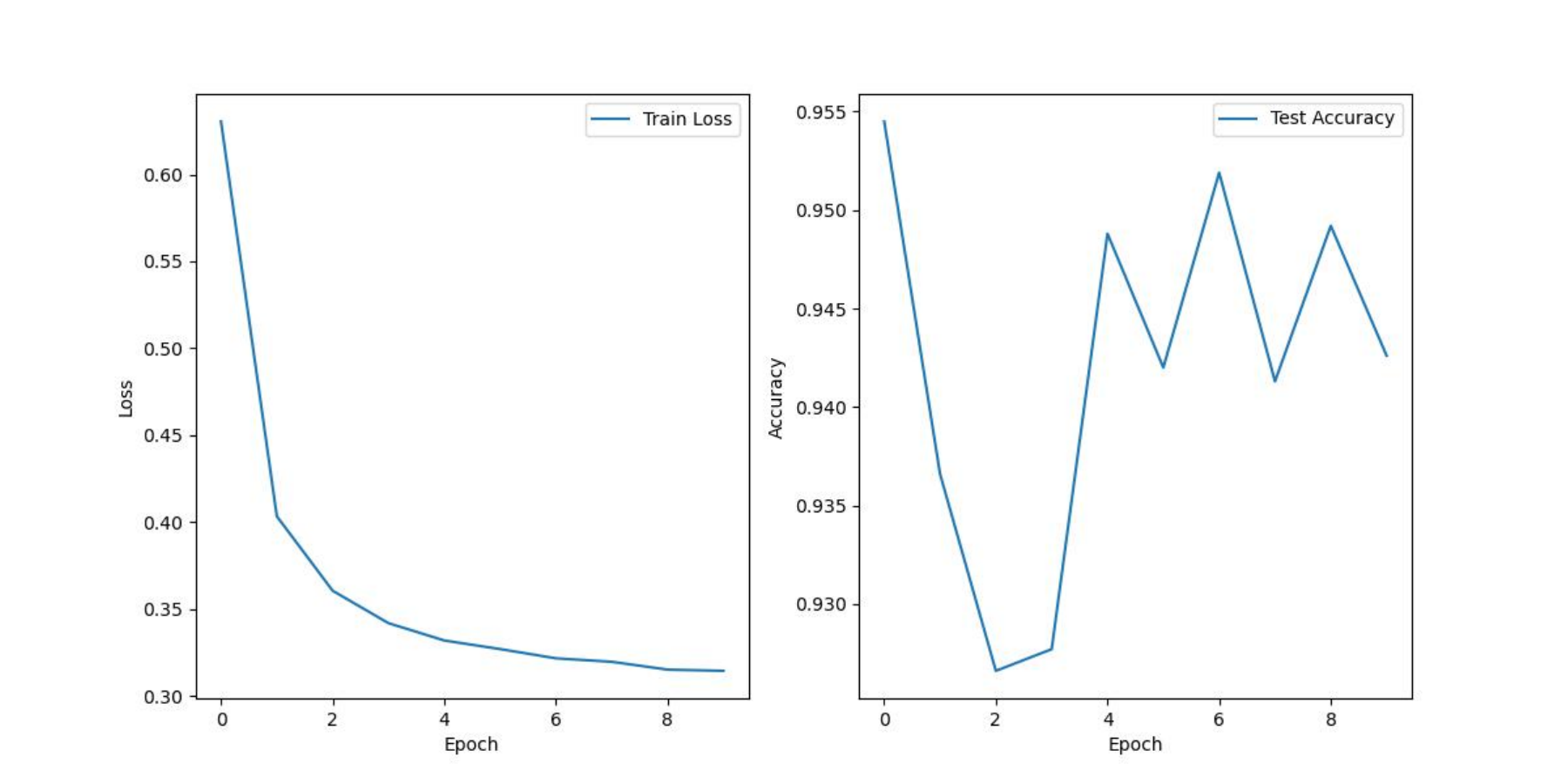}
    \caption{Learning curves during adversarial training. (Left) The training loss steadily decreases. (Right) The test accuracy on the original, clean dataset shows a slight initial dip before recovering, indicating a trade-off between standard accuracy and adversarial robustness.}
    \label{fig:training_curves}
\end{figure}

Next, we performed adversarial training on the baseline GAT decoder for 10 epochs using the methodology described in Sec. II.D. The training and testing curves are shown in Fig.~\ref{fig:training_curves}.

The test accuracy on the clean dataset slightly decreased from $\sim$96\% to $\sim$94.5\%, a common trade-off when improving model robustness. We then re-evaluated the original RL agent against this newly robust decoder. The ASR plummeted from 91.2\% to \textbf{16.2\%}.

\begin{figure}[H]
    \centering
    \includegraphics[width=0.9\columnwidth]{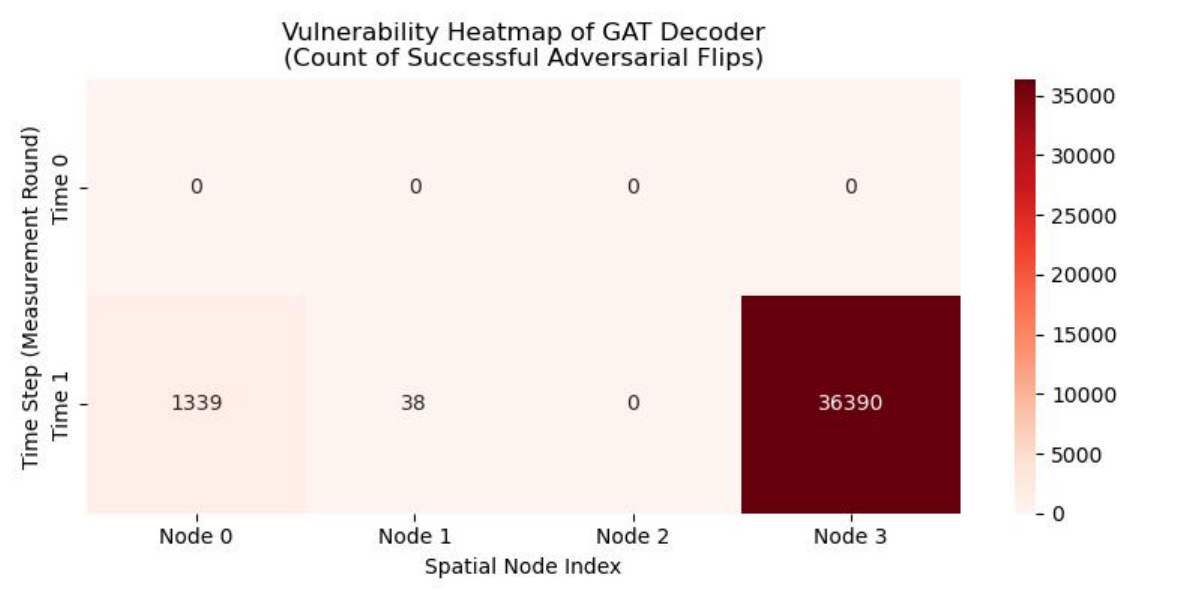}
    \caption{Vulnerability heatmap of the adversarially trained (robust) GAT decoder. The original vulnerability at (Node 0, Time 1) has been significantly mitigated. The RL agent adapted and found a new primary vulnerability at (Node 3, Time 1).}
    \label{fig:heatmap2}
\end{figure}

We then retrained a new RL agent from scratch against this robust decoder to see if it could discover new vulnerabilities. The new agent learned a different strategy, reflected in the new vulnerability heatmap in Fig.~\ref{fig:heatmap2}.

The results are clear: the vulnerability at `(Node 0, Time 1)` was successfully patched, with attack successes dropping by 96\% (from 36,610 to 1,339). However, the agent adapted and discovered a new, previously unexploited vulnerability at `(Node 3, Time 1)`, which now accounts for the majority of successful attacks.

\section{Discussion}
Our findings demonstrate that an RL agent can serve as a powerful and automated "red team" for auditing the reliability of neural network QEC decoders. The agent was not only able to find vulnerabilities but also to quantify their severity, discovering that the baseline decoder could be fooled with a high success rate by flipping just a single bit.

The specificity of the discovered vulnerability—concentrated at `Time 1`—suggests that the GAT decoder may have developed a heuristic that overweights the final syndrome measurement when making a decision. This is a subtle bias that would be difficult to identify through conventional performance metrics alone.

The shift in the vulnerability heatmap after adversarial training (Fig.~\ref{fig:heatmap1} vs. Fig.~\ref{fig:heatmap2}) vividly illustrates the "cat-and-mouse" dynamic of adversarial machine learning. While we successfully "vaccinated" the decoder against its most prominent weakness, the adversary adapted and found the next best attack vector. This highlights that achieving true robustness is not a single-step process but may require an iterative cycle of attack and defense. This iterative framework provides a clear path forward for systematically hardening decoders against a progressively wider range of failure modes.

\section{Conclusion and Future Work}
In this work, we have successfully demonstrated a novel framework for evaluating and enhancing the robustness of GNN-based QEC decoders using reinforcement learning. Our RL-based adversarial agent autonomously discovered critical, non-trivial vulnerabilities in a GAT decoder, which were subsequently mitigated through targeted adversarial training. This result validates the capability of GNNs to learn complex error features from data, but also underscores the importance of systematically probing for and defending against adversarial blind spots.

Future work will focus on iterating this adversarial training cycle. By repeatedly training an RL agent against the newly robust decoder and using the discovered exploits to further retrain the decoder, we aim to converge towards a model with no easily exploitable single- or few-bit-flip vulnerabilities. Furthermore, this methodology can be extended to different QEC codes and more realistic, hardware-specific noise models. Exploring more advanced RL algorithms, such as Actor-Critic (A2C) or Proximal Policy Optimization (PPO), could also lead to more efficient discovery of complex, multi-step attack strategies, further pushing the boundaries of decoder reliability.


\begin{acknowledgments}
This work was conducted as an independent project by the author.
\end{acknowledgments}

\end{document}